\documentclass[10pt,english,prb,amsmath,amssymb,prb,showkeys,superscriptaddress,twocolumn,showpacs,floatfix]{revtex4-2}
\usepackage{graphicx}
\usepackage{tabularx}
\usepackage{epstopdf}
\usepackage{dcolumn}
\usepackage[colorlinks=true,linkcolor=blue ,citecolor=blue,urlcolor=blue]{hyperref}
\usepackage{multirow}
\usepackage[usenames,dvipsnames]{xcolor}
\usepackage{soul}
\usepackage{braket}
\usepackage{bm}
\usepackage{nicefrac}
\usepackage{siunitx}
\usepackage{color,transparent}
\usepackage[utf8]{inputenc}
\usepackage{upgreek}

\usepackage[version=4]{mhchem}
\usepackage{chemformula}
\usepackage{amsmath}   
\usepackage{mathtools} 
\usepackage{amssymb}   
\usepackage{mathrsfs}  
\usepackage{textcomp}  
\usepackage{accents}   
\usepackage{bm}        
\usepackage{soul}
\usepackage{relsize}
\usepackage{graphicx}  
\usepackage{epstopdf}  
\usepackage{subcaption} 
\usepackage{caption}
\usepackage{makecell}
\usepackage{hhline}
\usepackage{cellspace}
\usepackage{array}     
\usepackage{tabularx}  
\usepackage{multirow}  
\usepackage{booktabs}  
\usepackage{dcolumn}   
\usepackage{gensymb}   
\usepackage{setspace}
\usepackage{scrextend}

\usepackage{xcolor}
\usepackage{soul}

\setstcolor{blue}
\begin{document}
\setlength{\heavyrulewidth}{0.08em}
\setlength{\lightrulewidth}{0.05em}
\setlength{\cmidrulewidth}{0.03em}
\setlength{\belowrulesep}{0.65ex}
\setlength{\belowbottomsep}{0.00pt}
\setlength{\aboverulesep}{0.40ex}
\setlength{\abovetopsep}{0.00pt}
\setlength{\cmidrulesep}{\doublerulesep}
\setlength{\cmidrulekern}{0.50em}
\setlength{\defaultaddspace}{0.50em}
\setlength{\tabcolsep}{4pt}

\title{Electron Polaron at Neutral 180\texorpdfstring{$^\circ$}{°} Domain Wall in PbTiO$_3$: \\ Stability, Trapping Energies, and Transverse Polarization}

\author{Mohammad Amirabbasi}
\email{amirabbasi@mm.tu-darmstadt.de}
\author{Jochen Rohrer}
\author{Karsten Albe}
\email{albe@mm.tu-darmstadt.de}

\affiliation{Technical University of Darmstadt, Materials Modelling Division, Otto-Berndt-Straße 3, Darmstadt D-64287, Germany}
\begin{abstract}
We use density-functional theory with a Hubbard correction to investigate Ti-centered electron polarons at neutral PbO-centered $180^\circ$ domain walls in tetragonal \ce{PbTiO3}. The Hubbard parameter for Ti $3d$ states is determined using 
the finite-size-corrected polaronic energy-level alignment  procedure, yielding stable electron-polaron formation in bulk PbTiO$_3$ with a trapping energy of $-$0.06 eV. 
In the domain-wall supercell, the excess electron localizes on Ti and forms a Ti$^{3+}$ center with an occupied $d_{xy}$ orbital in-gap state. Comparison of bulk-like and near-wall Ti sites shows that their trapping energies differ by only about 0.01 eV, indicating that this neutral domain wall does not provide a significant thermodynamic driving force for electron-polaron segregation. 
While the Ising-like reversal of the out-of-plane polarization is preserved, the localized electron induces a finite transverse polarization component normal to the wall, enhancing a local N\'eel-like distortion that is strongest when the polaron is located at the wall.
These results show that neutral $180^\circ$ domain walls in PbTiO$_3$ do not substantially alter the stability of Ti-centered electron polarons, but they can couple to the polaron-induced lattice distortion through a localized transverse polarization response.
\end{abstract}

\maketitle
\section{Introduction}

Ferroelectric perovskite oxides such as BaTiO$_3$, PbZr$_x$Ti$_{1-x}$O$_3$, and PbTiO$_3$ are widely studied for non-volatile memories, sensors, actuators, and capacitor technologies because their spontaneous polarization can be switched by external fields \cite{Scott_2007,Acosta2017,muralt2002pzt}. Below the Curie temperature, the paraelectric phase transforms into a polar state with several symmetry-equivalent polarization variants. The crystal therefore subdivides into domains separated by atomically thin domain walls (DWs), which reduce depolarizing fields and elastic energy \cite{Tagantsev_2010}. Because DWs can be created, displaced, and reconfigured by electric fields, they are increasingly viewed not only as passive microstructural features but also as functional nanoscale elements for domain-wall electronics \cite{salje2013domain,Meier2022}.

The functional response of ferroelectric DWs is strongly affected by defects and charge-compensation mechanisms. Dopants, vacancies, and excess carriers can modify DW mobility, pinning, local conductivity, and polarization structure; consequently, predictive control of ferroelectric properties requires a microscopic understanding of how charged species interact with different DW types \cite{Lee2011,Nataf2020,Chan2013,Bencan2020,Klomp2022,Marton2025}. This issue is particularly important because ferroelectric DWs may be either neutral, as in side-by-side or head-to-tail configurations, or charged, as in head-to-head and tail-to-tail configurations. Charged DWs carry bound polarization charge and therefore generate local electrostatic potentials that must be compensated \cite{Catalan2012,Meier_2015,Maksymovych2011,Tian-2010}. One possible compensation mechanism is the accumulation of free carriers at the wall \cite{Zuo2014,Sluka2012}. However, the generally low conductivity and high resistivity of many ferroelectric oxides indicate that compensation can also occur through localized valence changes of lattice ions or through self-trapped carriers \cite{Seidel2009,Bein2019,Rojac2017}.

Self-trapped carriers, or small polarons, form when an excess electron or hole localizes on an atomic site together with a local lattice distortion \cite{franchini2021polarons}. In oxide ferroelectrics, such localization is directly connected to reduction or oxidation of host ions and can therefore set charge-compensation limits and pin the Fermi level \cite{Klein2023}. Polarons are thus relevant not only for electronic and optical properties, but also for the response of domain walls to charged defects and non-stoichiometry \cite{Varley2012,faust1994free,ambrosio2018origin,sana2017,Schirmer_2006,Schirmer_2009,Vikhnin2002,Ambrosio2019}. Recent theoretical and experimental studies have shown that DWs can act as trapping sites for localized charges or charged defects, thereby influencing DW conductivity, pinning, and transport in ferroelectric and ferroelastic materials \cite{Bencan2020,Rojac2017,xu2024electronic,Korbel2018,Xiao2018,schroder2012,Meier2012,Eliseev2012,PRL2025}. These results motivate a direct comparison of polaron stability in bulk-like and DW environments.

PbTiO$_3$ provides a well-defined model system for this purpose.  At room temperature, PbTiO$_3$ adopts the tetragonal $P4mm$ phase after a paraelectric-to-ferroelectric transition at approximately 763 K \cite{Sicron_1994,Glazer_1978,Garcia_1996,Nehlmes_1990,Fontana_1990,Burns_1973}. Its electronic structure and lattice parameters have been extensively characterized, with a band gap of about 3.2--3.6 eV and an experimental tetragonality of $c/a \approx 1.06$ \cite{Bilc_2008,Schafranek2011,Benthem2001,Joseph2000,dmowski2001structure,Rodriguez_2002,Yamanaka-2018}. The structure and energetics of neutral $180^\circ$ DWs in PbTiO$_3$ have also been investigated previously, showing that PbO-centered walls are energetically favored over TiO$_2$-centered walls \cite{Meyer2002,Behera2011JPCM,Poykko1999Ab}. These properties make PbTiO$_3$ an appropriate reference material for isolating the interaction between a neutral ferroelectric DW and a localized excess electron.

\begin{figure*}
    \includegraphics[width=1.0\textwidth]{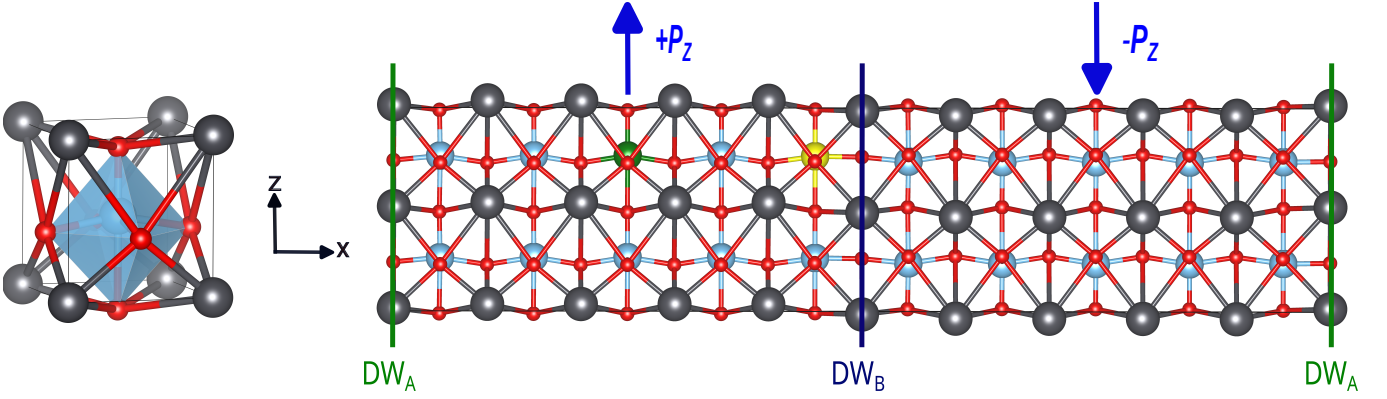}
    \caption{(Color online)  (left panel) Crystal structure of PbTiO$_{3}$ primitive cell for up polarization, plotted using VESTA~\cite{Momma_2011}. Ti ions are represented by cyan spheres, Pb ions by black spheres, and O ions by red spheres. Each Ti is surrounded by six oxygen atoms in this system to form a distorted octahedron. The down polarization state is achieved by the rotation of the up polarization state by 180 degrees around the \(x\)-axis. (right panel) The optimized 10$\times$2$\times$2 supercell size of 180$^\circ$ DW. The Green and dark-blue solid lines denote two symmetric DW. Notably, within the DW, the Ti-O bond lengths along the epical direction undergo variations. Similarly, positions of the Pb ions also exhibit changes across the DW. The green and yellow spheres denote the location of the installing electron polaron in bulk and wall regions relative to DW, respectively. \label{DW}
}
\end{figure*}

In bulk tetragonal PbTiO$_3$, Ti-centered electron polarons have been studied in detail and are known to correspond to the reduction of Ti$^{4+}$ to Ti$^{3+}$, with the excess electron localized in Ti $3d$ states \cite{Eglitis_2002,Ghorbani_2022,Windsor_2024}. In contrast, O-centered hole polarons are energetically unfavorable as shown by DFT calculations \cite{Erhart2014}. Despite this established bulk behavior, it remains unclear whether a neutral $180^\circ$ DW substantially changes the trapping energy of a Ti-centered electron polaron, and whether such a localized carrier modifies the local polarization structure of the wall. This distinction is important: if the DW lowers the polaron trapping energy, it may act as a segregation site for self-trapped electrons; if not, then the polaron remains essentially bulk-like and the neutral wall does not provide a thermodynamic driving force for charge localization.

In this work, we address this question using density functional theory with a Hubbard correction (DFT+$U$). We first determine an appropriate $U$ parameter for Ti $3d$ states using a finite-size-corrected level-alignment procedure, since the predicted stability of localized polarons is sensitive to self-interaction errors and finite-size effects~\cite{Cococcioni_2005, Falleta_2020, Falletta2022}. We then construct a PbO-centered neutral $180^\circ$ DW in tetragonal PbTiO$_3$ and compare Ti-centered electron polarons located in bulk-like and near-wall regions. By analyzing trapping energies, electronic density of states, and local polarization profiles, we determine whether the neutral DW promotes electron-polaron segregation and how the localized charge couples to the Ising-like polarization reversal at the wall.

\section{Computational Details}

\begin{figure}[!htbp]
  \centering
  \includegraphics[width=0.8\columnwidth]{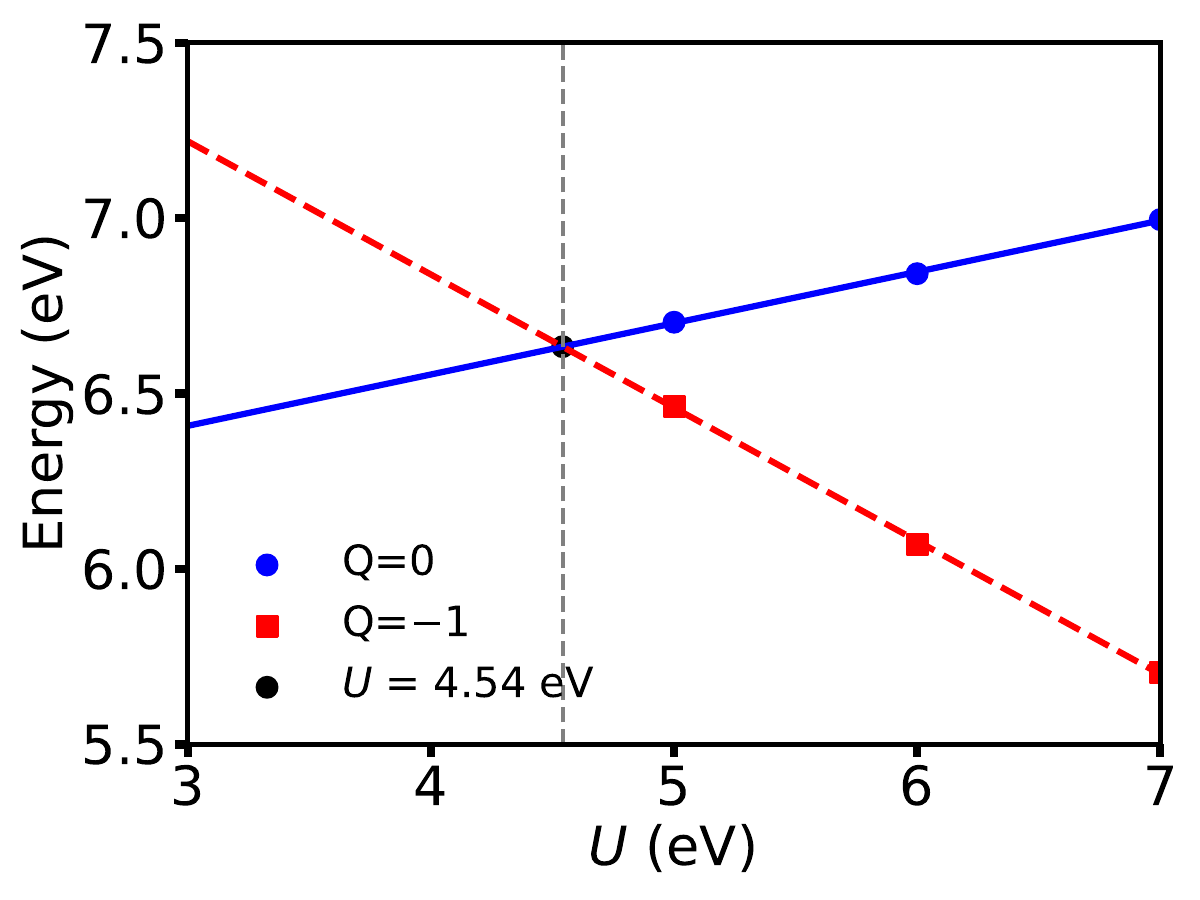}
  \caption{\label{fig:hubbard}
    Hubbard correction $U_{\mathrm{Ti}\text{-}3d}$ for a $3\times3\times3$ supercell with a polaronic state.
  }
\end{figure}

First-principles calculations were performed using the Vienna \textit{ab initio}
simulation package (VASP) \cite{KRESSE199615,Kresse_1999}. The electron--ion
interaction was described by the projector augmented-wave (PAW) method
\cite{Bloch_1994,Kresse_1999}, and exchange--correlation effects were treated
within the PBEsol form of the generalized-gradient approximation
\cite{Perdew_1996,Perdew_2008}. The valence configurations were
$6s^2 6p^2$ for Pb, $3s^2 3p^6 3d^2 4s^2$ for Ti, and $2s^2 2p^4$ for O.
Plane-wave cutoff energies of 900~eV and 520~eV were used for primitive-cell
and supercell calculations, respectively.

The tetragonal PbTiO$_3$ primitive cell was initialized from the experimental
structure \cite{Joseph2000}. Both lattice vectors and internal coordinates were
relaxed at the PBEsol level. Since the structural tetragonality of PbTiO$_3$ is
sensitive to the choice of exchange--correlation functional, and since adding a
Hubbard correction during the structural optimization reduces the PBEsol
$tetragonality$, the PBEsol-optimized volume was used as the reference geometry
for subsequent calculations. The Hubbard correction was applied  in
calculations involving excess-charge localization, where self-interaction errors
directly affect the stability of the polaronic state. Brillouin-zone sampling
was performed using $\Gamma$-centered meshes of $10\times10\times10$ for the
primitive cell and $3\times3\times3$ for the $3\times3\times3$ bulk supercell.
All ionic relaxations were continued until the residual forces were below
$5\times10^{-3}$ eV/\AA.

A neutral $180^\circ$ domain wall was constructed by joining oppositely
polarized tetragonal PbTiO$_3$ domains along the $[100]$ direction, with the
spontaneous polarization oriented along $[001]$ (see Fig. \ref{DW}). The downward-polarized domain
was generated by rotating the upward-polarized structure by $180^\circ$ about
the $[001]$ axis. We considered a PbO-centered wall, which is the lower-energy
configuration for neutral $180^\circ$ domain walls in tetragonal PbTiO$_3$
\cite{Meyer2002,Behera2011JPCM,Poykko1999Ab}. The pristine wall structure was
first relaxed in a $10\times1\times1$ supercell and then expanded to a
$10\times2\times2$ supercell for the polaron calculations. The corresponding
$\Gamma$-centered $k$-point meshes were $2\times8\times8$ and $2\times4\times4$,
respectively.

Localized electron polarons were generated by adding one excess electron to the
supercell and initially constraining the occupation matrix of a selected Ti
center. This procedure breaks the symmetry of the delocalized conduction-band
state and promotes localization of the excess electron on a chosen Ti site.
During the following ionic relaxation, the surrounding Ti--O bonds respond to
the localized charge and form the accompanying lattice distortion. After this
distortion was established, the occupation constraint was removed and the system
was fully re-relaxed self-consistently within DFT+$U$. Occupation-matrix control
was used to avoid convergence to metastable or delocalized solutions associated
with the multi-minimum character of DFT+$U$ calculations \cite{allen2014occupation}.

The Hubbard parameter applied to Ti $3d$ states was not treated as an empirical
fitting parameter. Instead, it was determined 
using the finite-size-corrected level-alignment procedure
of Falletta \textit{et al.} \cite{Falleta_2020}, in which the polaronic
Kohn--Sham level is aligned between the fully occupied and empty charge states
after removing spurious electrostatic finite-size contributions (see Fig. \ref{fig:hubbard})).
This point is important because the value
of $U$ controls the balance between electron localization and delocalization:
too small a $U$ leaves excessive self-interaction error and favors an extended
conduction-band state, whereas too large a $U$ can over-localize the excess
electron.

For the occupied polaron level, the electrostatic correction to the
Kohn--Sham eigenvalue is written as
\begin{equation}
    \epsilon_{\mathrm{corr}}^{\mathrm{occ}}
    =
    -\frac{2}{q} E_{\mathrm{m}}(q,\epsilon_0),
\end{equation}
whereas for the empty polaron level it is
\begin{equation}
    \epsilon_{\mathrm{corr}}^{\mathrm{unocc}}
    =
    -\frac{2}{q+q'_{\mathrm{pol}}}
    E_{\mathrm{m}}(q+q'_{\mathrm{pol}},\epsilon_\infty).
\end{equation}
Here, $q$ is the net charge of the simulation cell,
$q'_{\mathrm{pol}}=-q(1-\epsilon_\infty/\epsilon_0)$ is the polarization charge
associated with the ionic relaxation around the polaron, and
$E_{\mathrm{m}}$ is the image-charge correction evaluated using the Freysoldt
scheme \cite{Freysoldt2009,Freysoldt2014}. The static dielectric constant
$\epsilon_0$ screens the long-range electrostatic interaction of the occupied,
relaxed polaron, whereas the high-frequency dielectric constant
$\epsilon_\infty$ is appropriate for the unoccupied level, for which only the
electronic polarization responds. We used the experimental dielectric constants
$\epsilon_{\infty,a}=6.64$, $\epsilon_{\infty,c}=6.63$,
$\epsilon_{0,a}=106.9$, and $\epsilon_{0,c}=28.6$ \cite{Foster}. This procedure
yields $U_{\mathrm{Ti}\text{-}3d}=4.54$ eV for Ti $3d$ states.

\section{Results and Discussion}
\label{section_III}
\subsection{Geometry optimization}
We start from the geometry optimization of the primitive cell in the tetragonal phase. We use an experimental CIF file~\cite{Joseph2000} as the starting point and optimize both the cell volume and ionic positions. 
For the study of the tetragonal phase of FE materials like PbTiO$_{3}$, selecting an appropriate pseudopotential approach to approximate the electron-ion interaction and exchange-correlation functional is crucial to achieving a \( c/a \) ratio close to the experimental value. PBE functional overestimates the \( c/a \) ratio ( \( c/a \)=1.23~\cite{Ghorbani_2022}), and the addition of \( U \) does not significantly improve this discrepancy (\( c/a \)=1.005~\cite{Ghorbani_2022}), although the other lattice constants are closer to the experimental measurements ($a=b=$3.98~\AA~\cite{Ghorbani_2022}). 
In contrast, we find that the PBEsol functional yields a \( c/a \) ratio of 1.08, which decreases to 1.01 with the inclusion of \( U \). Given that the PBEsol functional provides a \( c/a \) ratio closer to the experimental value, we adopt the PBEsol-derived volume for subsequent calculations. 
In addition, the \( U \) is considered only for optimizing ionic positions in the following analysis when we consider excess charge for formation of polaron. The optimized lattice constants using PBEsol exchange-correlation functional are determined to be \(a = b = 3.87 \, \text{\AA}\) and \(c = 4.19 \), resulting in a \(c/a\) ratio of 1.08. These values show good agreement with the experimental data~\cite{Joseph2000, dmowski2001structure, Rodriguez_2002} (see Tab.~\ref{tab:Tab-symmetry}).
\begin{table}[!htbp]
\centering
\caption{Comparison of experimentally measured and theoretically calculated lattice constants and $c/a$ ratios for the tetragonal phase of PbTiO$_{3}$, highlighting data from various studies in the scientific literature.
}
\begin{tabular}{cccc}
\toprule
Methods&$a = b$ (\AA) &  $c$ (\AA) & $c/a$\\
\midrule
Exp.~\cite{dmowski2001structure, Yamanaka-2018}&3.90& 4.15&1.06\\
Exp.~\cite{Joseph2000}&3.90& 4.13&1.06\\
PBE~\cite{Ghorbani_2022}&3.84&4.72&1.23\\
PBE+$U$~\cite{Ghorbani_2022}&3.98&4.00&1.005\\
PBEsol+$U$  &3.94&3.99&1.01\\
PBEsol   &3.87&4.19&1.08\\
\bottomrule
\end{tabular}
\label{tab:Tab-symmetry}
\end{table}
\subsection{Electron-polaron trapping  in bulk phase}
Prior to investigating the interaction between the electron polaron and the DW, we  evaluated the trapping of the polaron on a bulk Ti site using supercells of various sizes.
The trapping energy of the electron polaron was evaluated using band-edge reference formulation \cite{Erhart2014, Wiktor2020},
\begin{equation}
    E_{\mathrm{trap}}
    =
    E_{\mathrm{pol}}^{-1}
    -
    E_{\mathrm{pristine}}^{0}
    -
    \epsilon_{\mathrm{CBM}}
    +
    E_{\mathrm{corr}},
    \label{eq:Etrap_cbm}
\end{equation}
where $E_{\mathrm{pristine}}^{0}$ is the energy of the neutral pristine
supercell and $\epsilon_{\mathrm{CBM}}$ is the conduction-band minimum of the
corresponding pristine system:
\begin{figure}[!b]
  {\includegraphics[width=0.8\linewidth]{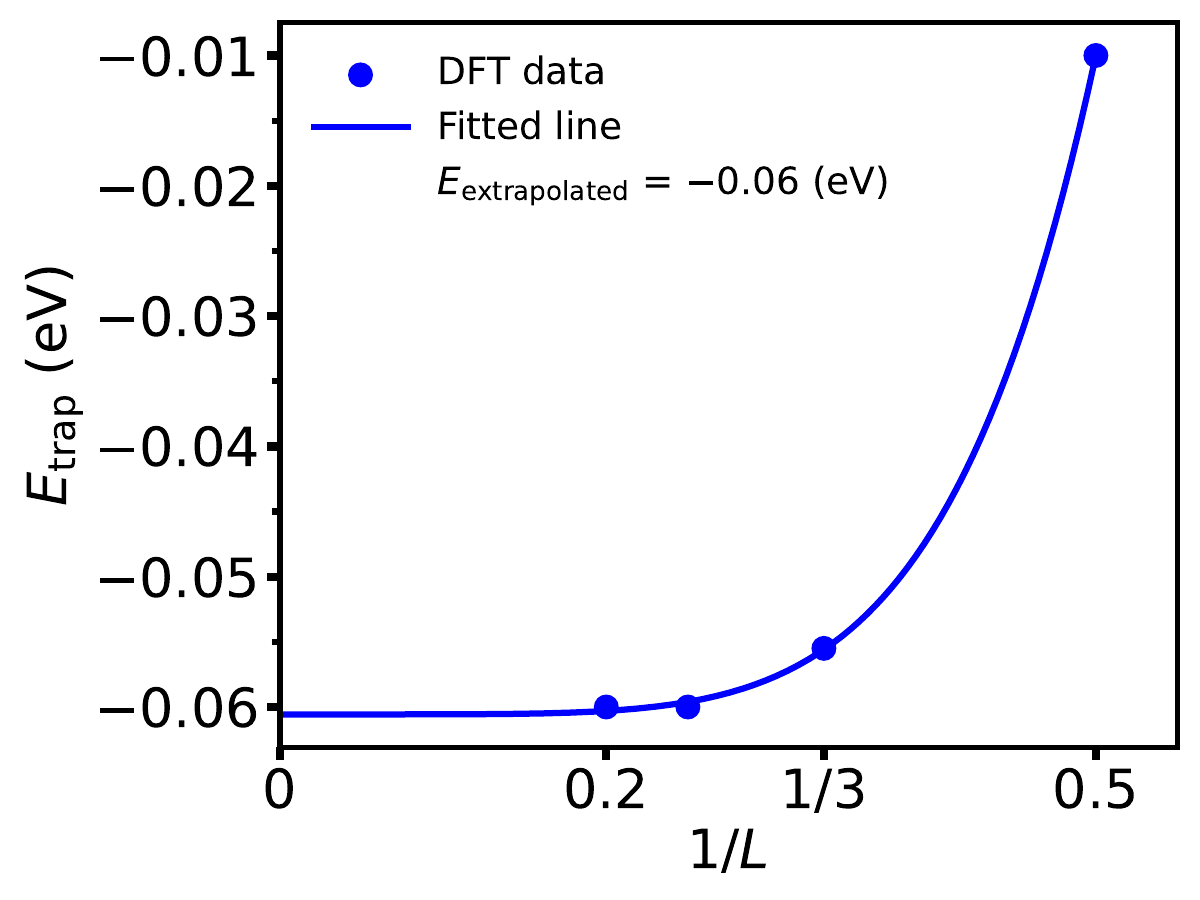}}
  \caption{\label{fig:trap} Polaron trapping energy as a function of inverse supercell size, (1/$L$), for $L$ $=$ 2–5.\label{fig:3-2}}
\end{figure}
In this formulation, the trapping energy converges systematically with supercell size, as shown in Fig.~\ref{fig:3-2}. The resulting finite-size-corrected trapping energy converges to $-$0.06 eV, differing from earlier reported values of $-$0.16 and $-$0.13 eV~\cite{Ghorbani_2022, Windsor_2024}. This discrepancy likely arises from differences in the $U$ parameter applied to Ti-3$d$ orbitals, the exchange-correlation functional, and dielectric constants, as well as the use of the delocalized electron configuration as the reference state in those studies.



\begin{figure}[!htbp]
  \centering
  \includegraphics[width=0.8\columnwidth]{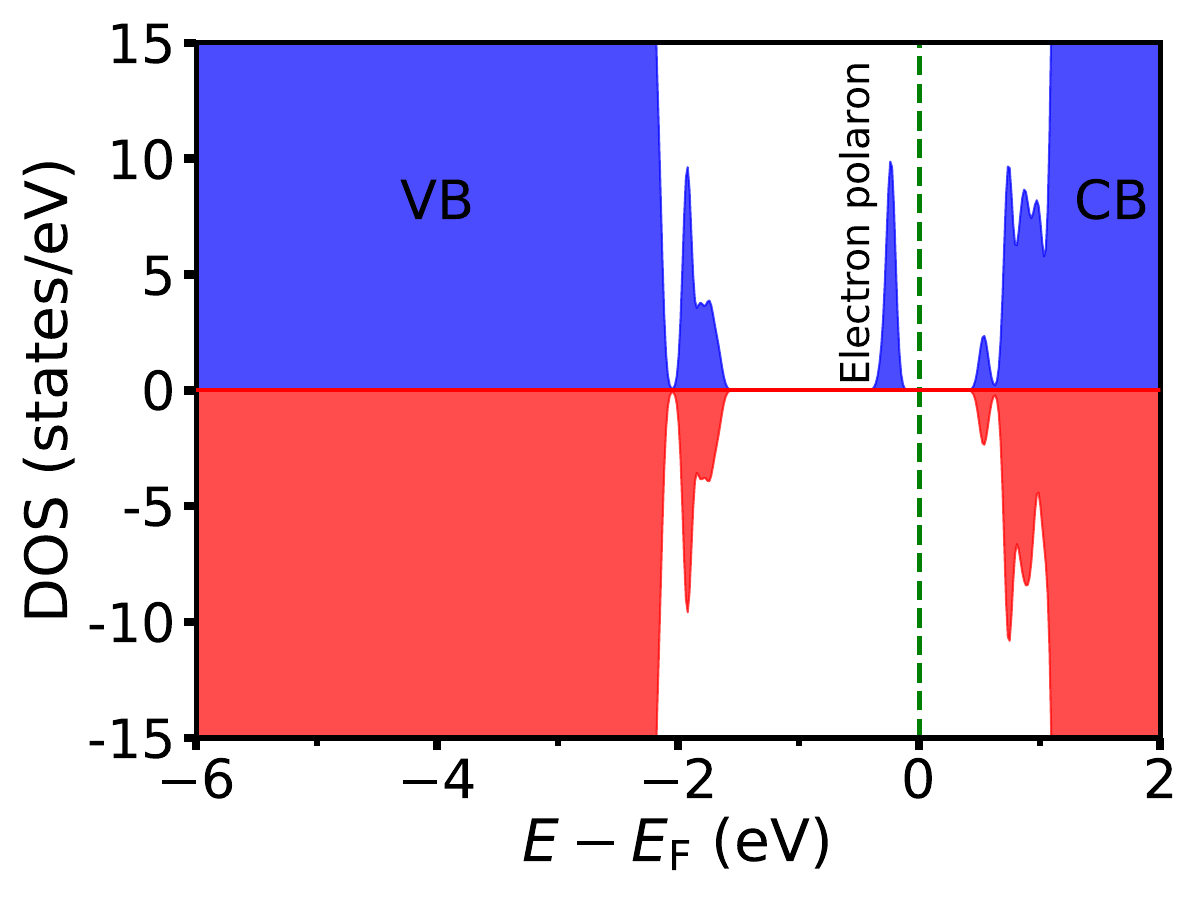}
  \caption{\label{fig:dos}
    DOS of the polaron configuration
    using the corrected $U_{\mathrm{Ti}\text{-}3d}$.
    Blue and red curves denote the spin-up and spin-down channels, respectively.
    The Fermi energy is set to zero, and the electron-polaron level lies
    0.75~eV below the CBM.
  }
\end{figure}

Figure~\ref{fig:dos}  displays the density of states (DOS) for the localized (polaronic) charge configurations. It reveals a distinct spin-up state below the Fermi energy, indicative of a stable localized polaron. Our calculations show that the additional electron predominantly occupies the \( d_\text{xy} \) orbital of the Ti ions, contributing a magnetic moment of \( 1 \, \mu_\text{B} \). Prior to polaron formation, the Ti ion is in the \( \text{Ti}^{4+} \) oxidation state, with an empty \( 3d \) shell. Upon localization, the ion is reduced to \( \text{Ti}^{3+} \), adopting a \( 3d^1 \) electronic configuration. The polaron level lies 0.75~eV below the CBM. 

\subsection{Electron-polaron on Ti-center: 180$^\circ$ Pb-center DW}

\begin{figure*}[!htp]
  \centering
  \begin{subfigure}{.32\linewidth}
    \centering
    \includegraphics[width=\linewidth]{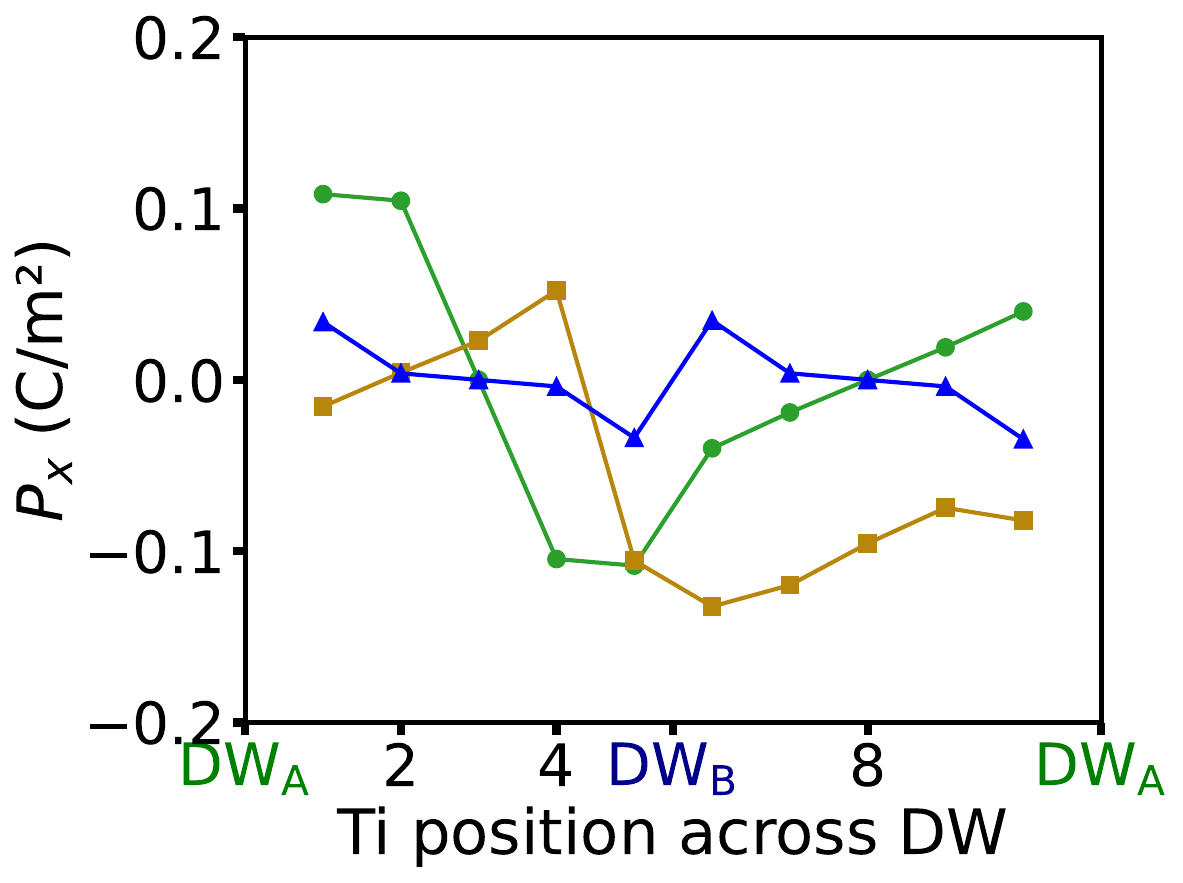}
    \caption{$x$-component polarization}
    \label{fig:5-1}
  \end{subfigure}
  \begin{subfigure}{.32\linewidth}
    \centering
    \includegraphics[width=\linewidth]{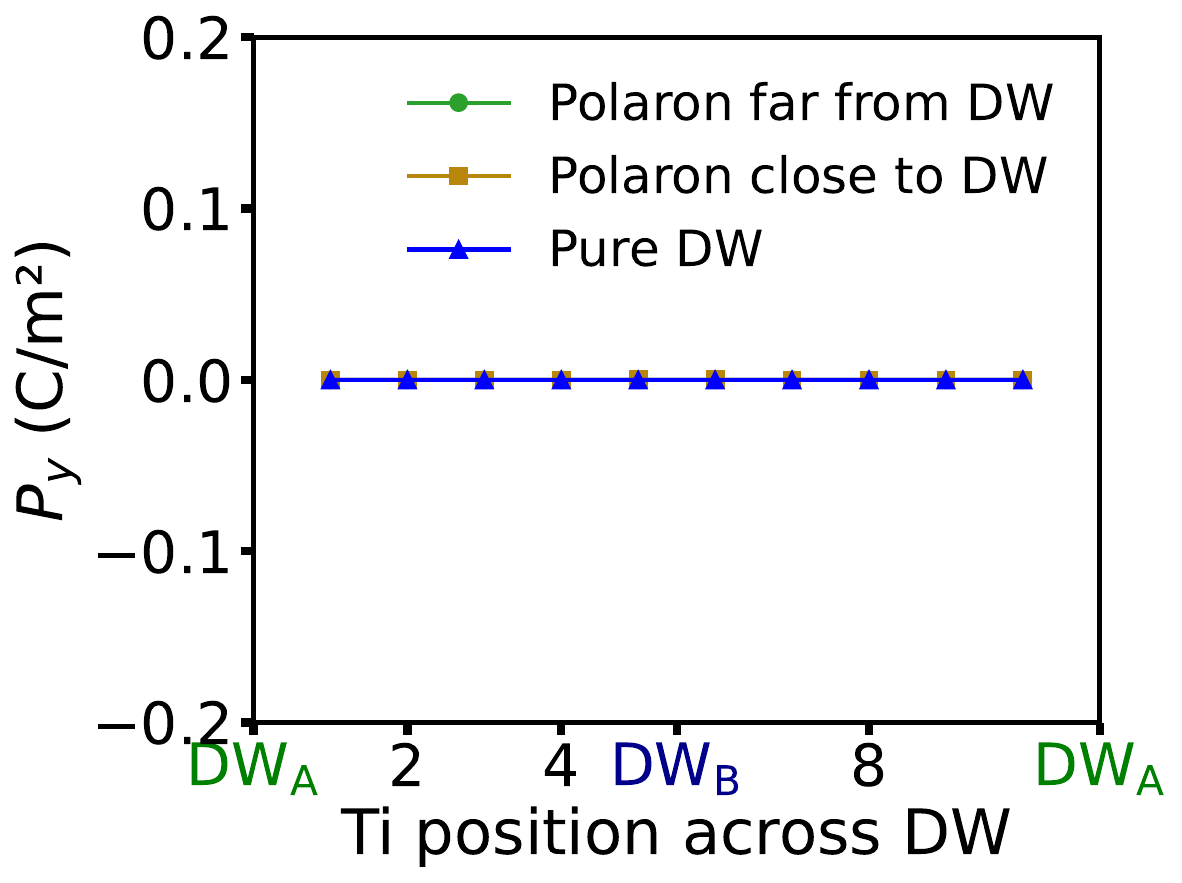}
    \caption{$y$-component polarization}
    \label{fig:5-2}
  \end{subfigure}
  \begin{subfigure}{.32\linewidth}
    \centering
    \includegraphics[width=\linewidth]{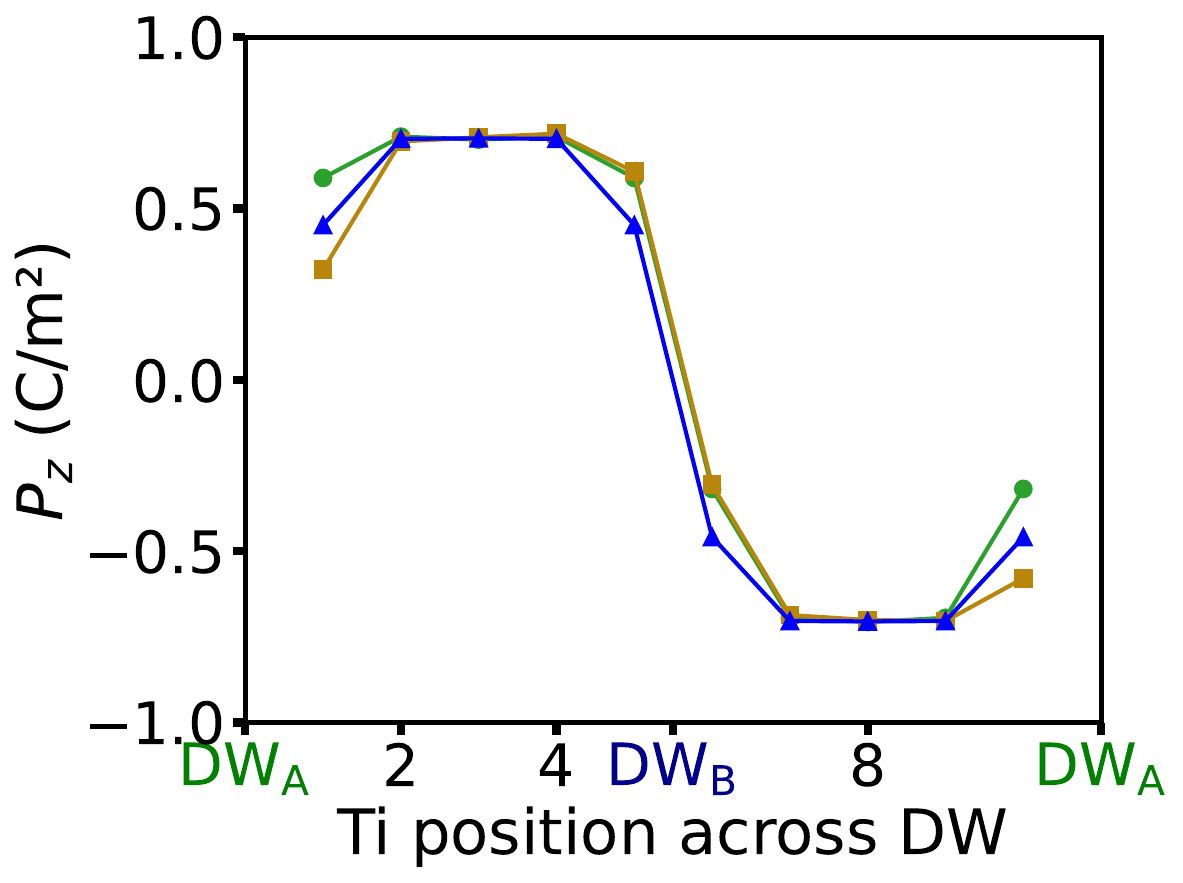}
    \caption{$z$-component polarization}
    \label{fig:5-3}
  \end{subfigure}
  \caption{(Color online)
    Polarization components across a neutral 180$^\circ$ Pb-centered domain wall in \ce{PbTiO3}.
    Panels show (a) $P_x$, (b) $P_y$, and (c) $P_z$ versus Ti position across the supercell. In the pure structure, $P_y$ remains negligible, while a small wall-normal N\'eel component, $P_x$, appears near the domain wall; $P_z$ dominates and reverses sign across the 180$^\circ$ wall. Adding a localized electron polaron preserves the $P_z$ and $P_y$ profiles, whereas a finite wall-normal component, $P_x$, develops at the polaron site.
  }
  \label{Fig-5}
\end{figure*}

We next examine the interaction between a Ti-centered electron polaron and a neutral
180$^\circ$ DW in tetragonal \ce{PbTiO3}. The calculations are performed
using a $10 \times 2 \times 2$ supercell containing 200 atoms using DFT+$U$. The supercell comprises
domains with opposite out-of-plane polarization, separated by a Pb-centered
180$^\circ$ DW, as shown in Fig.~\ref{DW}. To distinguish bulk-like and DW-related
polaron behavior, the excess electron is localized on Ti sites either far from or close
to the DW, indicated by the green and yellow spheres in Fig.~\ref{DW}, respectively.
The domain wall formation energy is calculated as
\begin{equation}
\gamma_\text{DW} = \frac{E_\text{DW}-E_\text{Bulk}}{2A},
\label{eq:DW-formation-energy}
\end{equation}
where $E_\text{DW}$ and $E_\text{Bulk}$ are total energies of the supercell structure containing DW and bulk, respectively. $A$ is the area of the domain wall. The resulting value of
\mbox{$\gamma_\mathrm{DW}=144~\mathrm{mJ\,m^{-2}}$},
is consistent with the result of previous first-principles calculations (128 and 132 $~\mathrm{mJ\,m^{-2}}$) for  180$^\circ$
domain walls in the Pb-O plane of tetragonal \ce{PbTiO3}~\cite{Behera2011JPCM, Meyer2002}.

To characterize the local polarization profile across the DW, we evaluate the
polarization associated with each Ti-centered unit using
\begin{equation}
    P_i =
    \frac{e}{\Omega}
    \sum_j
    \frac{q_j}{n_j}
    \left(
    r_i^j-r_i^{\mathrm{Ti}}
    \right),
    \label{eq:local_pol}
\end{equation}
where $\Omega$ is the unit-cell volume, $n_j$ is the multiplicity of atomic species
$j$, and $r_i^j-r_i^{\mathrm{Ti}}$ denotes the displacement of atom $j$ relative to the
central Ti ion along Cartesian direction $i$. The Born effective charges are taken
from Ref.~\onlinecite{Zhong1994}, with
$q_{\mathrm{Ti}}=+6.71$, $q_{\mathrm{Pb}}=+3.92$,
$q_{\mathrm{O_\mathrm{apical}}}=-5.51$, and
$q_{\mathrm{O_\mathrm{equatorial}}}=-2.56$.

The resulting polarization profiles are shown in Fig.~\ref{Fig-5}. In the pristine DW
structure, the polarization is dominated by the out-of-plane Ising component, $P_z$,
which changes sign abruptly across the wall. The corresponding DW width is confined
to approximately one primitive unit cell ($\sim\!3.87$~\AA), consistent with previous
first-principles calculations for Pb-centered 180$^\circ$ DWs in \ce{PbTiO3}
\cite{Meyer2002}. The transverse in-plane component, $P_y$, remains negligible
throughout the supercell. By contrast, a small wall-normal N\'eel-like component,
$P_x$, develops in the vicinity of the DW, in agreement with the behavior reported by
Behera \textit{et al.}~\cite{Behera2011JPCM}. Its maximum value in our calculation,
$P_x = 0.035~\mathrm{C\,m^{-2}}$, is close to the previously reported value of
$0.024~\mathrm{C\,m^{-2}}$. The calculated saturation value of the Ising component,
$P_z \approx 0.7~\mathrm{C\,m^{-2}}$, also agrees well with experimental values
($0.5$--$1.0~\mathrm{C\,m^{-2}}$)~\cite{lines2001principles,Abe2020} and with
zero-temperature first-principles predictions
($0.87$--$1.14~\mathrm{C\,m^{-2}}$)
\cite{TADMOR20022989,polar1998,Meyer2002,Kingslan2020,Zhang2017}.

Introducing a localized electron polaron on a Ti site leaves the overall Ising
polarization profile essentially unchanged. In particular, the reversal of $P_z$ across
the DW and the negligible transverse component $P_y$ are retained for polarons both
far from and close to the DW. The main structural response is instead a localized
enhancement of the wall-normal polarization component at the polaron site. For a
polaron located far from the DW, $P_x$ reaches
$0.11~\mathrm{C\,m^{-2}}$, whereas for a polaron located close to the DW the maximum
value is reduced to $0.05~\mathrm{C\,m^{-2}}$. Thus, the Ti-centered electron polaron
induces a local wall-normal polar distortion without significantly modifying the
underlying 180$^\circ$ Ising DW profile. This behavior is analogous to the local
transverse or wall-normal polar distortions previously reported for point defects such
as oxygen vacancies interacting with ferroelectric DWs~\cite{Petra2021,TOMODA2015}.

Finally, we evaluate the energetic stability of the Ti-centered electron polaron at two
representative positions, one in a bulk-like region and one close to the DW. The
occupied polaronic state is located $0.64$~eV below the conduction-band minimum
(CBM) for the bulk-like Ti site and $0.59$~eV below the CBM for the Ti site near the
DW. The corresponding trapping energies are $-0.04$~eV and $-0.03$~eV,
respectively, indicating that electron-polaron formation is favorable
in both cases.

The trapping energies, however, differ by only $0.01$~eV. This very small energy difference is
within the expected uncertainty associated with finite-size and self-interaction
effects in the elongated $10 \times 2 \times 2$ DW supercell. In particular, the
bulk-like trapping energy obtained in this geometry differs from the value calculated
for the more isotropic $3 \times 3 \times 3$ bulk supercell, reflecting residual
self-interaction of the localized excess electron in the DW supercell. We therefore
do not interpret the small difference between the bulk-like and near-DW trapping
energies as a meaningful site preference. Rather, these results show that the neutral
Pb-centered 180$^\circ$ DW provides no significant thermodynamic driving force for
Ti-centered electron-polaron segregation.

\section{Conclusion}
\label{section_IV}

In this work, we investigated the interaction of 
electron polarons with a 180$^\circ$ domain wall in \ce{PbTiO3}. While hole polarons on oxygen sites
are not expected for \ce{PbTiO3} \cite{Ghorbani_2022},  
electron localizes on Ti centers with a trapping energy of $-$0.06 eV. In the
domain-wall supercell, the Ti-centered electron polaron produces a local enhancement
of the wall-normal polarization component, while the main out-of-plane Ising
polarization profile remains essentially unchanged. The occupied polaronic level lies
below the conduction-band minimum, but the calculated trapping energies are small. In
particular, the trapping energies for a bulk-like Ti site and a Ti site close to the
domain wall differ by only $0.01$~eV.

This behavior contrasts with what has been reported for oxygen vacancies at a 180$^\circ$ domain wall in \ce{PbTiO3}. Oxygen vacancies are point defects that can couple more
directly to the local wall structure and have been shown to induce transverse or
wall-normal polarization components at domain walls \cite{hee2003}. In this respect, the
Ti-centered electron polaron produces a qualitatively similar local polarization
response: it also generates a finite wall-normal component, $P_x$, at the polaron
site. However, the energetic effect is much weaker for the isolated electron
polaron (10~meV vs 250~meV). While oxygen vacancies can act as stronger structural perturbations and may
promote defect accumulation or local wall modification, the electron polaron studied
here mainly induces a local polar distortion without appreciable segregation to the
neutral wall. 

\section*{Acknowledgement}
Authors acknowledge financial support from the Collaborative Research Center FLAIR (Fermi level engineering applied to oxide electroceramics), funded by the German Research Foundation (DFG) under Project-ID No. 463184206–SFB 1548 (project A02).
Lichtenberg and Paderborn Supercomputing Centers are gratefully acknowledged
by MA as the providers of needed computing facilities. The
guidelines provided by Prof. Julia Wiktor, Dr. Lorenzo Villa, Dr. Marcel Sadowski, Dr. Gustav Bhilmayer, Dr. Zhenbang Dai, Prof. Feliciano Giustino
 and Dr. Stefano Falletta are highly valued 
 by MA.


\bibliographystyle{apsrev4-2}
\bibliography{polaronDW.bib} 
\end{document}